\documentclass[10pt]{iopart}
\usepackage{graphicx,bm,amssymb,tabls,url}
\eqnobysec

\newcommand{\beq}{\begin{equation}}
\newcommand{\beqa}{\begin{eqnarray}}
\newcommand{\eeq}{\end{equation}}
\newcommand{\eeqa}{\end{eqnarray}}

\newcommand{\abs}[1]{\left\vert#1\right\vert}
\newcommand{\ket}[1]{\vert#1\rangle}
\newcommand{\braket}[2]{\langle#1\vert#2\rangle}
\newcommand{\braopket}[3]{\langle#1\vert#2\vert#3\rangle}
\newcommand{\dd}{{\rm d}}
\renewcommand{\det}{\mathop{\rm det}\,}
\newcommand{\frad}[2]{\displaystyle{\frac{#1}{#2}}}
\newcommand{\frat}[2]{\textstyle{\frac{#1}{#2}}}
\newcommand{\gue}{{\rm GUE}}
\newcommand{\h}{\widehat}
\newcommand{\half}{{\frac{1}{2}}}
\newcommand{\ii}{{\rm i}}
\newcommand{\lam}{\lambda}
\newcommand{\mean}[1]{{\langle#1\rangle}}
\renewcommand{\o}{\omega}
\newcommand{\osc}{{\rm osc}}
\renewcommand{\tr}{\mathop{\rm tr}\,}
\newcommand{\vx}{{\bm x}}
\newcommand{\vp}{{\bm p}}
\newcommand{\Ai}{{\rm Ai}}
\newcommand{\E}{{\rm even}}
\newcommand{\F}{{\rm F}}
\renewcommand{\H}{{\cal H}}
\newcommand{\Nabla}{{\bm\nabla}}
\renewcommand{\O}{{\rm odd}}
\newcommand{\1}{{\bm I}}

\newcommand{\xx}{\{x_n\}}
\newcommand{\uu}{\{u_n\}}
\newcommand{\yy}{\{y_n\}}
\newcommand{\yyuu}{\{y_n+u_n\}}
\newcommand{\pp}{\{p_n\}}
\newcommand{\zz}{\{z_n\}}
\newcommand{\vxx}{\{\vx_n\}}
\newcommand{\vpp}{\{\vp_n\}}

\begin{document}

\title{Expansion of a free Fermi gas released from an isotropic trapping potential}

\author{J M Luck$^1$ and P L Krapivsky$^{2,3}$}

\address{$^1$ Universit\'e Paris-Saclay, CNRS, CEA, Institut de Physique Th\'eorique, 91191~Gif-sur-Yvette, France}

\address{$^2$ Department of Physics, Boston University, Boston, MA 02215, USA}

\address{$^3$ Santa Fe Institute, Santa Fe, NM 87501, USA}

\begin{abstract}
We consider a system of non-interacting fermions prepared in the many-body ground state
of an isotropic trapping potential in any dimension,
and investigate the ballistic expansion of the fermionic cloud after the potential is suddenly released.
Using semi-classical techniques, we derive the full late-time profile of the expanding cloud
in the regime when the fermion number is large.
We thus obtain explicit expressions for power-law potentials with arbitrary exponent $a$ and in all dimensions $d$.
The momentum distribution and the spatial profile of the cloud exhibit
a universal edge exponent $d/a$,
thus generalizing the Wigner semi-circle law and the Thomas-Fermi distribution.
\end{abstract}

\eads{\mailto{jean-marc.luck@ipht.fr},\mailto{pkrapivsky@gmail.com}}

\maketitle

\section{Introduction}
\label{intro}

The discovery of Bose-Einstein condensates in 1995
triggered a considerable interest in the physics of ultracold atomic gases,
where statistics and interactions play a central part
(see~\cite{Bloch,Pethick,Pitaevskii} for overviews).
Ultracold Fermi gases quickly entered the scene and were shown to exhibit a rich physics
(see~\cite{Giorgini,Ketterle,Ketterle2} for reviews).
The key feature of fermionic gases is that one can take advantage of the so-called Feshbach resonances
in order to fine tune the scattering length measuring the strength of two-body interactions.
The crossover between a regime of superfluidity and a regime of dimer condensation can thus be explored.
The resonant or unitary Fermi gas, where the scattering length is formally infinite,
lies somewhere in the middle of this celebrated BCS-BEC crossover
(see~\cite{Strinati} for a recent review).

In most situations ultracold atomic gases are prepared in a restricted geometry,
either in a single confining potential well or in an optical lattice.
In the first case, if the trapping potential is suddenly released, the gas undergoes a phase of expansion.
If collisions after release are negligible,
the system enters a far-field regime at late times,
where the expansion is ballistic
and the shape of the atomic cloud maps out the momentum distribution of its initial state.
The case of a harmonic trapping potential has been investigated thoroughly.
In the simplest situation of a one-dimensional harmonic oscillator with time-dependent frequency $\o(t)$,
the dynamics is entirely characterised by a reduced length scale $\lam(t)$
obeying the non-linear second-order differential equation
\beq
\ddot\lam+\o(t)^2\lam=\frac{\o_0^2}{\lam^3},
\label{erma}
\eeq
known as the Ermakov-Pinney equation~\cite{ermakov,pinney}
(see~\cite{coelho} for a recent overview).
In the case of an instantaneously released trap,
this yields $\lam(t)=\sqrt{1+(\o_0t)^2}$ (see~(\ref{lamfree})),
which exhibits the ballistic growth $\lam(t)\approx\o_0t$.
The Ermakov-Pinney equation has been rediscovered countless times,
and extended to a wealth of harmonically trapped many-body systems
in various dimensions~\cite{Popov,CD,Kagan,Kagan2,Bruun,Gleis,Menotti,Castin,Minguzzi,MinguzziG,Vicari}.
This approach always yields a ballistic, possibly anisotropic, spreading of the fermionic cloud.
In more recent years, an approach referred to as generalized hydrodynamics
has been proposed to describe free expansion in one-dimensional integrable models,
among other non-equilibrium and transport properties
(see~\cite{BHK} for an overview).
Finally, we mention a very recent work~\cite{MDK}, also based on a hydrodynamic approach,
suggesting the possibility of a sub-ballistic spreading of some interacting quantum gases in one and two dimensions.

In the present work we investigate at depth the expansion of a non-interacting Fermi gas
following the instantaneous release of a trapping potential.
The main focus is on the limit shape of the ballistically expanding cloud,
for which we derive analytical results for arbitrary isotropic confining potentials in any dimension.
This paper is a follow-up to an earlier work by our group~\cite{KLM}
on the late-time behavior of the return probability or Loschmidt echo of one-dimensional trapped fermionic gases.
The advantage of free fermionic systems is that they lend themselves to analytical treatments,
thus giving access to detailed information which would otherwise be unavailable.
We consider especially the regime where the fermion number $N$ is large,
so that most of the occupied one-body orbitals are highly excited,
allowing for the use of semi-classical methods.
In the case of isotropic power-law potentials,
the expressions thus obtained appear as generalizations of the Wigner semi-circle law.
To our knowledge these outcomes are novel, in spite of the relative simplicity of the situation at hand.

The detailed setup of this paper and its main results are as follows.
The one-dimensional case is investigated at depth in section~\ref{1d}.
For a harmonic potential (section~\ref{1dhar}), the expansion dynamics can be studied analytically at all times.
The positions of the fermions are given by $x_n(t)=z_n\lam(t)/\sqrt\o$ (see~(\ref{xzsca})),
where $\lam(t)$ obeys the Ermakov-Pinney equation~(\ref{erma}),
whereas the $z_n$ are distributed as the eigenvalues of a GUE matrix of size $N$ (see~(\ref{GUE})).
This exact similarity solution corroborates the picture of a late-time far-field regime.
This line of thought is put forward for an arbitrary trapping potential in section~\ref{1dlate},
predicting a ballistic expansion of the form $x_n\approx p_nt$,
where the momenta $p_n$ can be viewed as $N$ independent classical random variables, with distribution (see~(\ref{sgal}))
\beq
f_N(p)=\sum_{n=0}^{N-1}\abs{\h\psi_n(p)}^2.
\label{insum}
\eeq
If the fermion number $N$ is very large, the use of semi-classical methods results in the general expression (see~(\ref{fnsemi}))
\beq
f_N(p)\approx\frac{1}{\pi}\int_{p^2/2}^{E_\F}\frac{\dd E}{V'(x)}\qquad(\abs{p}<p_\F),
\label{inf}
\eeq
where $x\ge0$ depends on $p$ and $E$ through $p^2=2(E-V(x))$, so that $x=0$ at the Fermi energy $E_\F=p_\F^2/2$.
In section~\ref{1dpower} we treat power-law potentials, $V(x)\sim\abs{x}^a$, with arbitrary exponent $a>0$,
and show that the momentum distribution of a large fermionic cloud reads (see~(\ref{fpower}))
\beq
f_N(p)\sim\frac{N}{p_\F}\left(1-\frac{p^2}{p_\F^2}\right)^{1/a}\qquad(\abs{p}<p_\F),
\eeq
so that the late-time profile of a large fermionic cloud reads
\beq
\rho_N(x;t)\sim\frac{N}{p_\F t}\!\left(\!1-\frac{x^2}{(p_\F t)^2}\!\right)^{1/a}\qquad(\abs{x}<p_\F t).
\eeq
The above expressions generalize the Wigner semi-circle law to all power-law potentials.
The Fermi momentum grows itself as $p_F\sim N^{a/(a+2)}$ (see~(\ref{1def})).
More specific results are presented in section~\ref{1dsquare} in the case of a square-well potential,
where the fermions are initially confined in an interval of length $L$.

We turn to the higher-dimensional situation in section~\ref{hid}.
For an arbitrary isotropic (i.e., radial) potential,
the late-time regime is again characterized by a ballistic expansion of the form $\vx_n\approx\vp_nt$,
where the momenta $\vp_n$ are drawn from a distribution of the form~(\ref{insum}),
where the sum runs over the $N$ initially occupied one-body states (see~(\ref{higal})).
Because of spin and orbital degeneracies,
there are in general several many-body ground states.
For some specific fermion numbers~$N$,
there is however a unique isotropic ground state consisting of complete shells of degenerate eigenstates.
If the fermion number $N$ is very large, the momentum distribution can again be estimated
by means of semi-classical methods (section~\ref{hilate}).
This yields the isotropic expression~(\ref{hisemi}), generalising~(\ref{inf}).
The case of a power-law potential with arbitrary exponent $a$ is dealt with in section~\ref{hipower}.
The momentum distribution of a large fermionic cloud reads (see~(\ref{hifpower}))
\beq
f_N(\vp)\sim\frac{N}{p_\F^d}\left(1-\frac{p^2}{p_\F^2}\right)^{d/a}\qquad(p=\abs{\vp}<p_\F),
\eeq
so that the late-time profile of a large fermionic cloud reads
\beq
\rho_N(r;t)\sim\frac{N}{(p_\F t)^d}\left(1-\frac{r^2}{(p_\F t)^2}\right)^{d/a}\qquad(r<p_\F t).
\eeq
The above isotropic expressions can be viewed as generalizations
of the Wigner semi-circle law and the Thomas-Fermi distribution to all power-law potentials in all dimensions,
the latter distributions being respectively recovered for $(d,a)=(1,2)$ and~(3,2).
The Fermi momentum grows itself as $p_F\sim N^{a/((a+2)d)}$ (see~(\ref{hief})).
Section~\ref{hidisk} contains more specific results on the two-dimensional example of an infinite circular well,
where the fermions are initially confined in a disk with unit radius.
A brief discussion of our findings is given in section~\ref{disc}.
An appendix is devoted to the derivation of the formulas~(\ref{fnwell}).

\section{The one-dimensional case}
\label{1d}

\subsection{General setting}
\label{1dgal}

We consider $N$ identical non-interacting spinless fermions in a trapping potential in one dimension.
The one-body Hamiltonian reads
\beq
\H=\frac{p^2}{2}+V(x),
\eeq
with $p=-\ii\,\dd/\dd x$.
For simplicity, the fermion mass $m$ and the Planck constant $\hbar$ are set equal to unity.
The potential $V(x)$ is even, and growing at infinity,
so that the spectrum of $\H$ is discrete.
Assuming for simplicity that the latter is non-degenerate,
we denote its eigenvalues by $E_n$ ($n=0,1,\dots$),
and by $\psi_n(x)$ the associated real normalized wavefunctions,
obeying\footnote{Here and throughout the following, accents denote derivatives.}
\beq
-\half\psi_n''(x)+V(x)\psi_n(x)=E_n\psi_n(x)
\eeq
and
\beq
\int_{-\infty}^\infty\psi_m(x)\psi_n(x)\,\dd x=\delta_{mn}.
\eeq

The system is prepared in its many-body ground state.
The fermions therefore occupy the $N$ lowest eigenstates of $\H$ ($n=0,\dots,N-1$).
The many-body wavefunction is given by a Slater determinant,
both in position space and in momentum space:\footnote{In this work,
all determinants are of size $N\times N$, with indices in the range $1\le k,l\le N$.}
\beqa
\braket{\Psi(0)}{\xx}&=&\frac{1}{\sqrt{N!}}\,\det(\psi_{k-1}(x_l)),
\nonumber\\
\braket{\Psi(0)}{\pp}&=&\frac{1}{\sqrt{N!}}\,\det(\h\psi_{k-1}(p_l)).
\eeqa
The Fourier transform, defined by the formulas
\beq
\h\psi(p)=\int_{-\infty}^\infty\psi(x)\,\e^{-\ii px}\,\frac{\dd x}{\sqrt{2\pi}},\quad
\psi(x)=\int_{-\infty}^\infty\h\psi(p)\,\e^{\ii px}\,\frac{\dd p}{\sqrt{2\pi}},
\eeq
is such that the $L^2$ norm of $\psi$ is the same in position and momentum space.

The trapping potential $V(x)$ is released instantaneously at $t=0$.
The ensuing expansion of the fermionic cloud is more conveniently described in momentum space.
We have
\beq
\braket{\Psi(t)}{\pp}=\prod_{n=1}^N\e^{-\half\ii t p_n^2}\braket{\Psi(0)}{\pp},
\eeq
and so
\beqa
\braket{\Psi(t)}{\xx}
&=&\prod_{n=1}^N\int_{-\infty}^\infty\e^{-\half\ii t p_n^2+\ii p_nx_n}\,\frac{\dd p_n}{\sqrt{2\pi}}
\,\braket{\Psi(0)}{\pp}.
\nonumber\\
\label{phit}
\eeqa

Hereafter we focus our attention onto the time dependence of simple one-body observables,
namely the total fermion density,
\beq
\rho_N(x;t)=\prod_{n=1}^N\int_{-\infty}^\infty\dd x_n\sum_{m=1}^N\delta(x-x_m)\,\abs{\braket{\Psi(t)}{\xx}}^2,
\label{rhodef}
\eeq
and the corresponding moments,
\beq
\mu_{2k}(t)
=\sum_{n=1}^N\braopket{\Psi(t)}{x_n^{2k}}{\Psi(t)}
=\int_{-\infty}^\infty x^{2k}\rho_N(x;t)\,\dd x,
\label{mudef}
\eeq
and especially on their asymptotic form at late times.

\subsection{Harmonic potential}
\label{1dhar}

In order to get some intuition regarding how to deal with the late-time regime in general,
it is useful to first consider an exactly solvable example, namely that of a harmonic potential,
\beq
V(x)=\frac{\o^2x^2}{2}.
\eeq
As recalled in the Introduction,
many-body quantum dynamics in time-dependent harmonic traps have been the subject of a good deal of investigations.
In the present setting of non-interacting fermions,
the harmonic case is the only example where the full time dependence
of the many-body wavefunction can be worked out explicitly.

The one-body energy levels read
\beq
E_n=\left(n+\half\right)\o\qquad(n=0,1,\dots).
\eeq
The corresponding wavefunctions have essentially the same form
in position space and in momentum space, namely
\beqa
\psi_n(x)&=&\frac{(\o/\pi)^{1/4}}{\sqrt{2^nn!}}\,H_n(x\sqrt{\o})\,\e^{-\o x^2/2},
\nonumber\\
\h\psi_n(p)&=&\frac{(-\ii)^n}{(\pi\o)^{1/4}\sqrt{2^nn!}}\,H_n(p/\sqrt{\o})\,\e^{-p^2/(2\o)},
\eeqa
where
\beq
H_n(z)=\sum_{k=0}^{\lfloor{n/2}\rfloor}(-1)^k\frac{n!}{k!(n-2k)!}(2z)^{n-2k}
\eeq
are the Hermite polynomials~\cite{Abramowitz}.
We thus have
\beq
\braket{\Psi(0)}{\pp}=
\frac{(-\ii/\sqrt{2})^{N(N-1)/2}}{(\pi\o)^{N/4}\sqrt{G(N+2)}}
\prod_{n=1}^N\e^{-\half z_n^2}\,\det(H_{k-1}(z_l)),
\label{phiph}
\eeq
with the notation
\beq
z_n=\frac{p_n}{\sqrt{\o}},
\eeq
and where
\beq
G(N+2)=\prod_{n=1}^N n!
\eeq
is the Barnes $G$ function or superfactorial.
The determinant entering~(\ref{phiph}) can be evaluated by subtracting
from each row a suitably chosen linear combination of the previous ones,
so that each Hermite polynomial $H_{k-1}(z_l)$ is eventually replaced by its leading term $(2z_l)^{k-1}$.
We thus obtain
\beq
\det(H_{k-1}(z_l))=2^{N(N-1)/2}\Delta(\zz),
\label{hvdm}
\eeq
where
\beq
\Delta(\zz)=\det(z_l^{k-1})=\prod_{1\le i<j\le N}(z_i-z_j)
\eeq
is a Vandermonde determinant.

Inserting~(\ref{phiph}) and~(\ref{hvdm}) into~(\ref{phit}), we get
\beqa
\braket{\Psi(t)}{\xx}
&=&\frac{(\o/\pi)^{N/4}(-\ii\sqrt{2})^{N(N-1)/2}}{\sqrt{G(N+2)}}
\nonumber\\
&\times&\prod_{n=1}^N\int_{-\infty}^\infty\e^{-\half(1+\ii\o t)z_n^2+\ii\sqrt{\o}z_nx_n}\,\Delta(\zz)\,\frac{\dd z_n}{\sqrt{2\pi}}.
\label{phix1}
\eeqa
Then, translating the integration variables according to
\beq
z_n=y_n+u_n,\qquad
u_n=-\frac{\ii\sqrt{\o}}{1+\ii\o t}\,x_n,
\label{zudef}
\eeq
the formula~(\ref{phix1}) becomes
\beqa
\braket{\Psi(t)}{\xx}
&=&\frac{(\o/\pi)^{N/4}(-\ii\sqrt{2})^{N(N-1)/2}}{\sqrt{G(N+2)}}
\,\exp\left(-\frac{\o}{2(1+\ii\o t)}\sum_{n=1}^Nx_n^2\right)
\nonumber\\
&\times&\prod_{n=1}^N\int_{-\infty}^\infty\e^{-\half(1+\ii\o t)y_n^2}\,\Delta(\yyuu)\,\frac{\dd y_n}{\sqrt{2\pi}}.
\label{phix2}
\eeqa
Expanding the Vandermonde determinant by rows,
choosing either $y_n$ or $u_n$ for each label $n$,
we obtain $2^N$ terms, all of which vanish upon integration over the variables~$\yy$,
except the constant term $\Delta(\uu)$.
The multiple integral entering~(\ref{phix2}) therefore equals $(1+\ii\o t)^{-N/2}\Delta(\uu)$.
Finally, taking the squared modulus of~(\ref{phix2}), and using the definition~(\ref{zudef}) of $u_n$, we get
\beqa
\abs{\braket{\Psi(t)}{\xx}}^2
&=&\frac{(2b(t)^2)^{N^2/2}}{(2\pi)^{N/2}G(N+2)}
\nonumber\\
&\times&\exp\left(-b(t)^2\sum_{n=1}^Nx_n^2\right)
\abs{\Delta(\xx)}^2,
\label{phixres}
\eeqa
with
\beq
b(t)=\sqrt\frac{\o}{1+(\o t)^2}=\frac{\sqrt{\o}}{\lam(t)},
\label{bdef}
\eeq
where the reduced length scale
\beq
\lam(t)=\sqrt{1+(\o t)^2}
\label{lamfree}
\eeq
obeys the Ermakov-Pinney equation~(\ref{erma}).

The result~(\ref{phixres}) describes the expansion of the fermionic cloud
at all times and for all values of the fermion number~$N$.
In other words, setting
\beq
x_n(t)=\frac{z_n}{b(t)}=\frac{z_n\,\lam(t)}{\sqrt{\o}},
\label{xzsca}
\eeq
the rescaled positions $z_n$ can be viewed as classical random variables,
whose joint distribution coincides with that
of the eigenvalues of a random Gaussian unitary $N\times N$ matrix $M$
in the standard GUE ensemble, namely (see~\cite{Mehta} for a full account)
\beqa
\rho(\zz)
&=&\frac{2^{N^2/2}}{(2\pi)^{N/2}G(N+2)}
\exp\left(-\sum_{n=1}^Nz_n^2\right)\abs{\Delta(\zz)}^2.
\label{GUE}
\eeqa
There are indeed profound relationships between determinantal processes, free fermions, and random matrix theory
(see~\cite{Johansson,Forrester} for overviews).
This analogy has been used at depth to investigate the statics of trapped non-interacting fermions,
either in their ground state or at finite temperature
(see~\cite{Lacroix,Dean} and the references therein).

\begin{figure}[!htbp]
\begin{center}
\includegraphics[angle=0,width=.65\linewidth,clip=true]{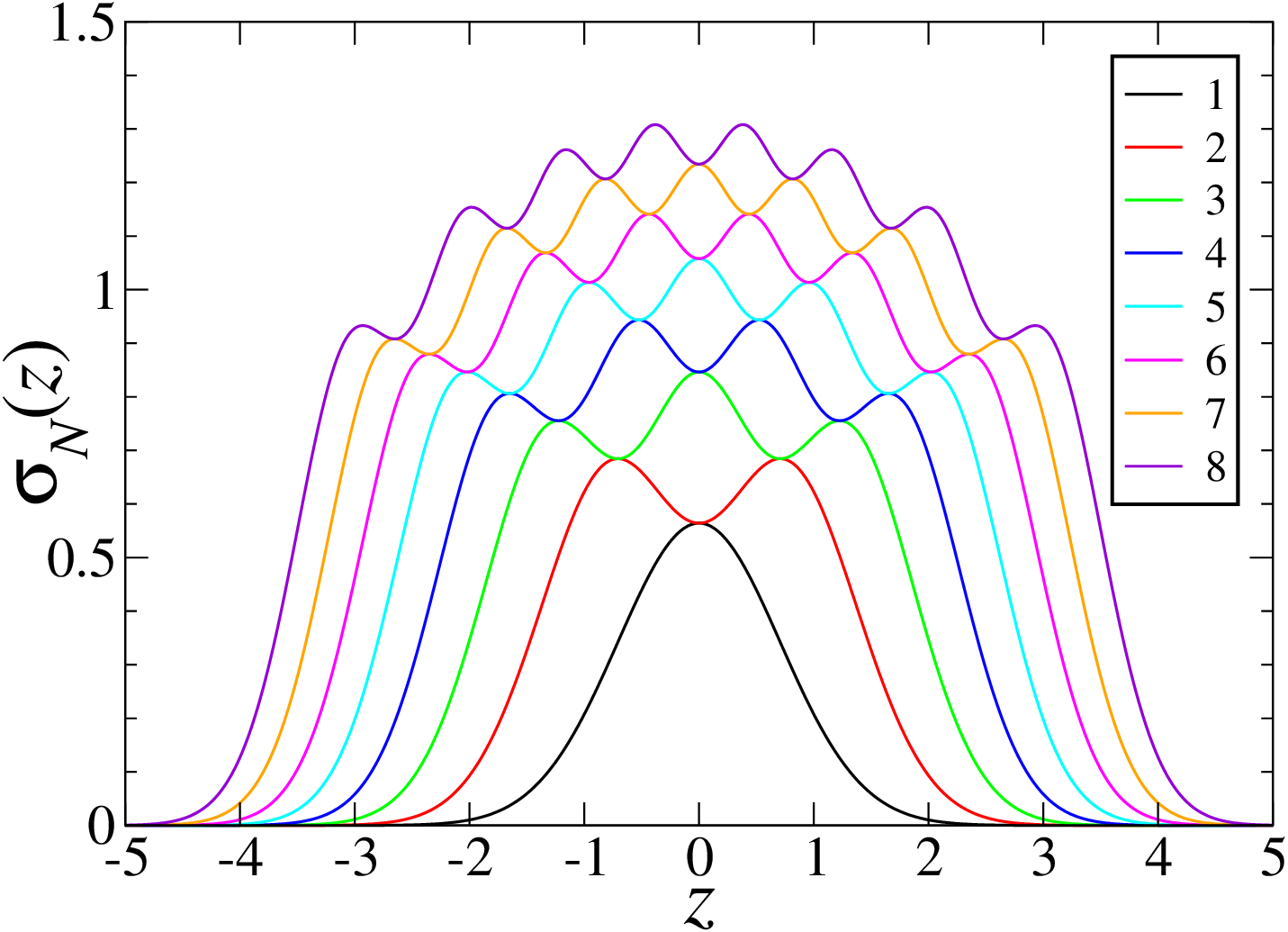}
\caption{
GUE eigenvalue density $\sigma_N(z)$ given by~(\ref{sig1}) plotted against $z$, for $N=1$ to 8 (see legend).}
\label{sigman}
\end{center}
\end{figure}

As a consequence of~(\ref{xzsca}), the particle density defined in~(\ref{rhodef}) reads
\beq
\rho_N(x;t)=b(t)\,\sigma_N(b(t)x),
\label{rhohar}
\eeq
where
\beq
\sigma_N(z)=\sum_{n=1}^N\mean{\delta(z-z_n)}=\mean{\tr\delta(z-M)}_\gue
\eeq
is the GUE eigenvalue density.
This quantity reads (see e.g.~\cite{Mehta})
\beq
\sigma_N(z)=\frac{\e^{-z^2}}{\sqrt{\pi}}\sum_{n=0}^{N-1}\frac{H_n(z)^2}{2^nn!}.
\label{sig1}
\eeq
Figure~\ref{sigman} shows how successive individual levels contribute to build up this quantity.

The sum entering~(\ref{sig1}) can be performed by means of the Christoffel-Darboux formula.
The outcome appears in several guises in the literature, including
\beqa
\sigma_N(z)
&=&\frac{\e^{-z^2}}{\sqrt{\pi}\,2^N(N-1)!}\left(H_N'(z)H_{N-1}(z)-H_{N-1}'(z)H_N(z)\right)
\nonumber\\
&=&\frac{\e^{-z^2}}{\sqrt{\pi}\,2^N(N-1)!}\left(H_N(z)^2-H_{N-1}(z)H_{N+1}(z)\right).
\label{sig2}
\eeqa
We mention for further reference that~(\ref{sig1}) can be recast as
\beq
\sigma_N(z)=\frac{1}{\sqrt{\o}}\sum_{n=0}^{N-1}\abs{\h\psi_n(\sqrt{\o}\,z)}^2
=\frac{1}{\sqrt{\o}}\,f_N(\sqrt{\o}\,z),
\label{siggal}
\eeq
and is therefore equivalent to~(\ref{sgal}).

As a consequence of~(\ref{rhohar}),
the moments of the positions of all fermions, defined in~(\ref{mudef}),
exactly evaluate to
\beq
\mu_{2k}(t)=\frac{\nu_{2k}(N)}{b(t)^{2k}},
\label{moms}
\eeq
where the dependences on time $t$ and on the fermion number $N$ factorize, with
\beq
\nu_{2k}(N)=\int_{-\infty}^{\infty}\sigma_N(z)\,z^{2k}\,\dd z=\mean{\tr M^{2k}}_\gue.
\eeq
These moments of the GUE distribution~(\ref{GUE}) can be evaluated by Wick's theorem.
The non-trivial underlying combinatorial problem has been solved in~\cite{Harer}, yielding
\beq
\nu_{2k}(N)=(2k-1)!!\sum_{l=0}^k2^{-(k-l)}{k\choose l}{N\choose l+1}.
\label{harer}
\eeq
The first few of these moments read
\beqa
&&\nu_0(N)=N,
\quad
\nu_2(N)=\frac{N^2}{2},
\quad
\nu_4(N)=\frac{N}{4}\,(2N^2+1),
\nonumber\\
&&\nu_6(N)=\frac{5N^2}{8}\,(N^2+2),
\quad
\nu_8(N)=\frac{7N}{16}\,(2N^4+10N^2+3).
\eeqa

Henceforth we consider the regime of late times $(\o t\gg1)$.
We have $b(t)\approx1/(\sqrt{\o}\,t)$, and so
\beq
x_n\approx z_n\sqrt{\o}\,t,
\eeq
where the $z_n$ have the GUE distribution~(\ref{GUE}).
This demonstrates the ballistic spreading of the fermionic cloud at late times.
The particle density scales as
\beq
\rho_N(x;t)\approx\frac{1}{\sqrt{\o}\,t}\ \sigma_N\!\left(\frac{x}{\sqrt{\o}\,t}\right),
\label{rhosca}
\eeq
and the corresponding moments grow as
\beq
\mu_{2k}(t)\approx\nu_{2k}(N)(\o t^2)^k.
\eeq

\begin{figure}[!htbp]
\begin{center}
\includegraphics[angle=0,width=.65\linewidth,clip=true]{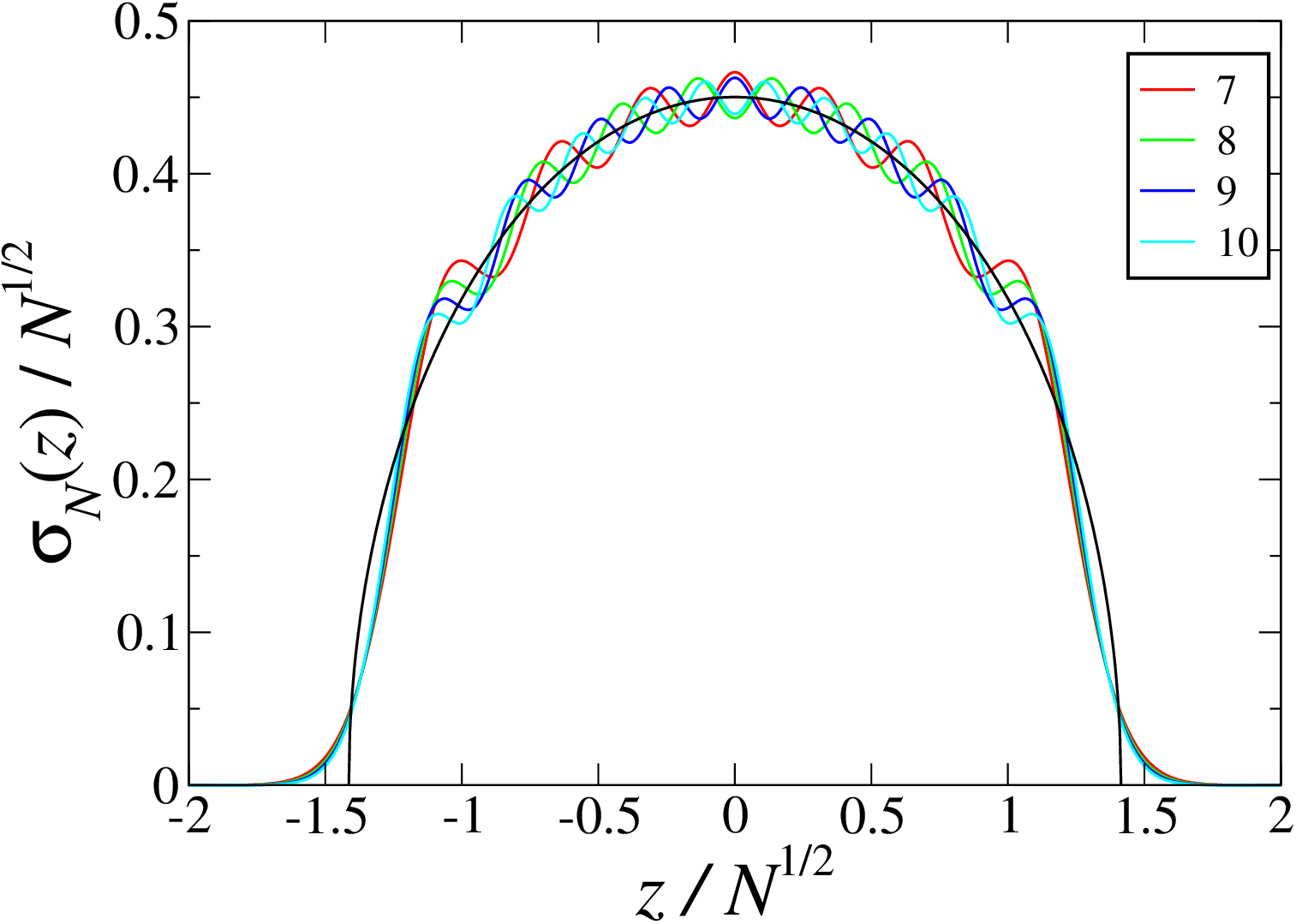}
\caption{
Scaling plot of $\sigma_N(z)/\sqrt{N}$ against $z/\sqrt{N}$ for $N=7$ to 10 (see legend).
Black curve: reduced Wigner semi-circle law ($y=\sqrt{2-x^2}/\pi$).}
\label{scasigma}
\end{center}
\end{figure}

When the fermion number $N$ is very large,
the eigenvalue density becomes the Wigner semi-circle law~\cite{Wigner},
\beq
\sigma_N(z)\approx\frac{\sqrt{2N-z^2}}{\pi}
\qquad(\abs{z}<\sqrt{2N}).
\label{sigmawigner}
\eeq
Its moments are dominated by the last term $(l=k)$ in~(\ref{harer}), namely
\beq
\nu_{2k}(N)\approx\frac{(2k-1)!!}{(k+1)!}\,N^{k+1}.
\eeq
The above results translate to
\beq
\rho_N(x;t)\approx\frac{\sqrt{2N\o t^2-x^2}}{\pi\o t^2}
\label{rhowigner}
\eeq
and
\beq
\mu_{2k}(t)\approx\frac{(2k-1)!!}{(k+1)!}\,N(N\o t^2)^k.
\eeq
The fermionic cloud is essentially supported by an interval expanding ballistically according to
\beq
\abs{x}<\sqrt{2N\o}\,t.
\eeq
Figure~\ref{scasigma} shows a scaling plot of $\sigma_N(z)/\sqrt{N}$ against $z/\sqrt{N}$,
exhibiting the convergence to the Wigner semi-circle law~(\ref{sigmawigner}).

\begin{figure}[!htbp]
\begin{center}
\includegraphics[angle=0,width=.65\linewidth,clip=true]{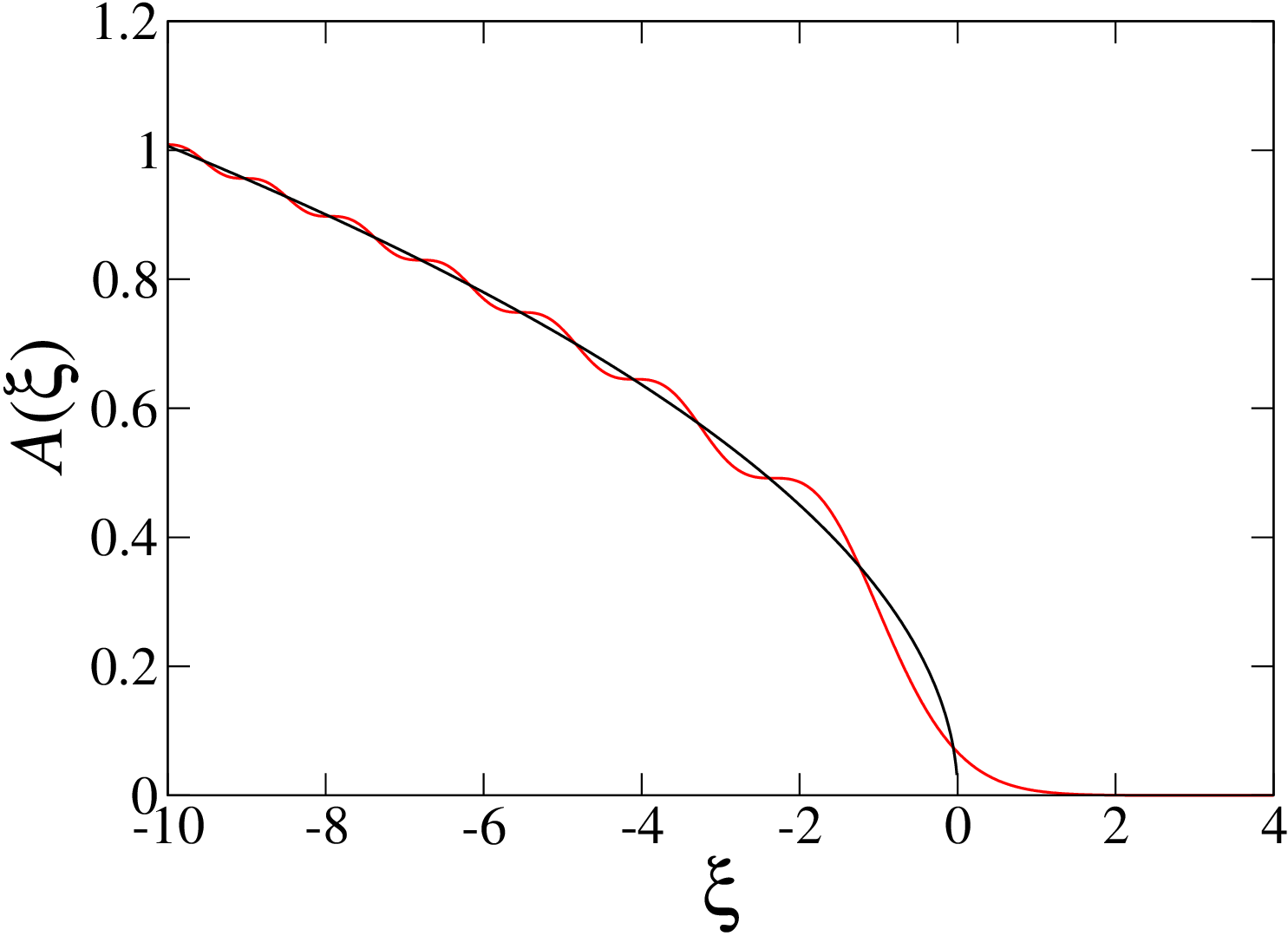}
\caption{
Function $A(\xi)$ entering the scaling law~(\ref{sigmaairy}).
Black curve: leading term of the asymptotic expansion~(\ref{ainf}),
matching the Wigner semi-circle law~(\ref{sigmawigner}).}
\label{Aplot}
\end{center}
\end{figure}

The leading finite-$N$ corrections to the Wigner semi-circle law are also known,
both in the bulk of the spectrum and near its edges~\cite{Forrester1,Kalisch,Garoni}.
In the bulk, setting
\beq
z=\sqrt{2N}\,\cos\theta\qquad(0<\theta<\pi),
\eeq
we have
\beq
\sigma_N(z)=\frac{\sqrt{2N}\,\sin\theta}{\pi}
\left(1-\frac{\cos[N(\sin 2\theta-2\theta)]}{4N\,\sin^3\theta}+\cdots\right).
\eeq
The leading term is nothing but~(\ref{sigmawigner}),
whereas the first correction is rapidly oscillating and proportional to $1/N$.
Near the edges of the spectrum, considering the upper edge for definiteness and setting
\beq
z=\sqrt{2N}\left(1+\frac{\xi}{(2N)^{2/3}}\right),
\eeq
we have
\beq
\sigma_N(z)\approx(2N)^{1/6}\,A(\xi),
\label{sigmaairy}
\eeq
with
\beq
A(\xi)=\Ai'(\xi)^2-\xi\,\Ai(\xi)^2,
\label{Adef}
\eeq
where $\Ai$ denotes the Airy function.
The sharp edge of the Wigner semi-circle law is thus broadened
into a smooth crossover whose relative extent shrinks as $N^{-2/3}$.
The following asymptotic expansions have been derived in~\cite{Forrester1}.
For $\xi\to-\infty$, towards the bulk of the spectrum,
we have
\beq
A(\xi)=\frac{\sqrt{\abs{\xi}}}{\pi}-\frac{\cos\left(\frat43\abs{\xi}^{3/2}\right)}{4\pi\abs{\xi}}+\cdots
\label{ainf}
\eeq
The leading term matches the semi-circle law~(\ref{sigmawigner}),
whereas the next term is oscillating (see~figure~\ref{Aplot}).
For $\xi\to+\infty$, away from the spectrum,
$A(\xi)$ falls off very fast, as
\beq
A(\xi)\approx\frac{17}{96\pi\sqrt\xi}\,\exp\left(-\frac43\xi^{3/2}\right).
\eeq
In particular, the mean number of fermions outside the Wigner interval has a finite limit value, namely
\beq
N_{\rm out}=2\int_0^\infty A(\xi)\,\dd\xi=\frac{1}{3\pi\sqrt3}\approx0.061258.
\eeq
This number is not only microscopic, but numerically very small~\cite{Johnstone}.

\subsection{Late-time regime in the general case}
\label{1dlate}

Let us come back to the setting of section~\ref{1dgal}, with an arbitrary even trapping potential~$V(x)$.
In the regime of late times,
it is legitimate to apply the stationary-phase approximation to the expression~(\ref{phit})
of the time-dependent many-body wavefunction.
We thus obtain the ballistic expansion law
\beq
x_n\approx p_n\,t,
\label{xbal}
\eeq
which is expected on physical grounds.
Furthermore, relative distances between distinct fermions also grow linearly in time,
so that the fermionic cloud is deep in the far-field regime, where interferences become negligible.

We thus arrive to the following picture of the late-time
regime of the expansion dynamics.
The positions of the fermions grow ballistically according to~(\ref{xbal}),
where the momenta $p_n$ can be viewed as $N$ independent classical random variables, with distribution
\beq
f_N(p)=\sum_{n=0}^{N-1}\abs{\h\psi_n(p)}^2,
\label{sgal}
\eeq
normalized as
\beq
\int_{-\infty}^\infty f_N(p)\,\dd p=N.
\eeq
The formula~(\ref{sgal}) is equivalent to~(\ref{siggal}) in the case of a harmonic potential.
The detailed analysis of this exactly solvable example performed in section~\ref{1dhar}
therefore corroborates the above derivation of~(\ref{xbal}),~(\ref{sgal}).

When the fermion number $N$ is very large,
most of the one-body states involved in~(\ref{sgal}) are highly excited $(n\gg1)$.
The distribution $f_N(p)$ can therefore be evaluated by semi-classical or WKB (Wentzel-Kramers-Brillouin) methods
(see e.g.~\cite{Berry,Eckhardt}, and~\cite{LL,Froman,Brack} for overviews).

Consider first a fixed large enough energy $E$.
The curve in the phase plane $(x,p)$ defined by the classical equation $\H(p,x)=E$ is even in both coordinates
and can be parametrised as
\beq
p(x;E)=\pm\sqrt{2(E-V(x))}.
\label{pofx}
\eeq
The branch with the $+$ sign is a decreasing function of $\abs{x}$,
from $p=\sqrt{2E}$ at $x=0$ to $p=0$ at both turning points $x=\pm x_0(E)$, such that $V(x)=E$.
The ensuing microcanonical distribution of $p$ is such that
\beq
f(p;E)\,\dd p=\frac{\dd t}{2T(E)},
\eeq
where
\beq
T(E)=\int_0^{x_0(E)}\frac{\dd x}{\sqrt{2(E-V(x))}}
\label{tedef}
\eeq
is the quarter-period of the classical motion at energy $E$.
Using Newton's equation
\beq
\frac{\dd p}{\dd t}=-V'(x),
\eeq
we obtain
\beq
f(p;E)=\frac{1}{2T(E)V'(x)}\qquad(\abs{p}<\sqrt{2E}),
\label{fpe}
\eeq
where $x\ge0$ depends on $p$ and $E$ through $p^2=2(E-V(x))$ (see~(\ref{pofx})).

The expression~(\ref{fpe}) can be used to approximate the momentum distribution as
\beq
f_N(p)\approx\sum_{n=0}^{N-1}f(p;E_n).
\label{fnsum}
\eeq
We introduce for further reference the Fermi momentum $p_\F$ and energy $E_\F$ by setting
\beq
E_\F=\frac{p_\F^2}{2}=E_N.
\eeq
As already said, most energy levels $E_n$ are highly excited.
They are therefore approximately given by the semi-classical quantization rule
\beq
S(E_n)\approx\frac{n\pi}{2},
\label{semi}
\eeq
where
\beq
S(E)=\int_0^{x_0(E)}\sqrt{2(E-V(x))}\,\dd x
\label{sedef}
\eeq
is the action of a quarter-period.
Viewing~(\ref{semi}) as a relationship between continuous variables $E$ and $n$, we have
\beq
\frac{\dd n}{\dd E}\approx\frac{2}{\pi}\,S'(E)=\frac{2}{\pi}\,T(E).
\label{neid}
\eeq
Inserting~(\ref{fpe}) in~(\ref{fnsum}),
and transforming the sum into an integral over~$n$,
and then over $E$ by means of~(\ref{neid}),
we are left with the following semi-classical estimate
\beq
f_N(p)\approx\frac{1}{\pi}\int_{p^2/2}^{E_\F}\frac{\dd E}{V'(x)}\qquad(\abs{p}<p_\F)
\label{fnsemi}
\eeq
for the momentum distribution of the $N$ fermions.
It is again understood that $x\ge0$ depends on $p$ and $E$ through $p^2=2(E-V(x))$.
Using the ballistic scaling~(\ref{xbal}),
the formula~(\ref{fnsemi}) provides a general quantitative prediction
for the profile of a large fermionic cloud at late times, namely
\beq
\rho_N(x;t)\approx\frac{1}{t}\,f_N\!\left(\frac{x}{t}\right).
\eeq
The cloud is therefore essentially supported by an interval expanding ballistically according to $\abs{x}<p_\F t$.

\subsection{Power-law potential}
\label{1dpower}

Let us now give the explicit forms of the above general results
in the case of a power-law trapping potential of the form
\beq
V(x)=\left(\frac{\abs{x}}{\ell}\right)^a,
\label{vpower}
\eeq
where $a>0$ is an arbitrary exponent, and $\ell$ is a measure of the trap size.

The positive turning point is $x_0(E)=\ell\,E^{1/a}$,
and so the expression~(\ref{tedef}) for the quarter-period evaluates to
\beqa
T(E)
&=&\int_0^{\ell\,E^{1/a}}\frac{\dd x}{\sqrt{2(E-(x/\ell)^a)}}
\nonumber\\
&=&\sqrt{\frac{\pi}{2}}\,\frac{\Gamma(1+1/a)}{\Gamma(1/2+1/a)}\,\ell\,E^{1/a-1/2}.
\eeqa
The expression~(\ref{fpe}) for the microcanonical momentum distribution therefore reads
\beq
f(p;E)=\frac{\Gamma(1/2+1/a)}{\sqrt{2\pi E}\;\Gamma(1/a)}\left(1-\frac{p^2}{2E}\right)^{-1+1/a}.
\eeq
Similarly, the expression~(\ref{sedef}) for the action of a quarter-period evaluates to
\beqa
S(E)
&=&\int_0^{\ell\,E^{1/a}}\sqrt{2(E-(x/\ell)^a)}\,\dd x
\nonumber\\
&=&\sqrt{\frac{\pi}{2}}\,\frac{\Gamma(1+1/a)}{\Gamma(3/2+1/a)}\,\ell\,E^{1/a+1/2}.
\eeqa
The semi-classical quantization formula~(\ref{semi}) therefore reads
\beq
E_n\approx\left(\sqrt{\frac{\pi}{2}}\,\frac{\Gamma(3/2+1/a)}{\Gamma(1+1/a)}\,\frac{n}{\ell}\right)^{\frat{2a}{a+2}},
\eeq
and so
\beq
E_\F=\frac{p_\F^2}{2}\approx\left(\sqrt{\frac{\pi}{2}}\,\frac{\Gamma(3/2+1/a)}{\Gamma(1+1/a)}\,\frac{N}{\ell}\right)^{\frat{2a}{a+2}}.
\label{1def}
\eeq
The estimate~(\ref{fnsemi}) for the momentum distribution evaluates to
\beq
f_N(p)\approx\frac{\Gamma(3/2+1/a)}{\sqrt\pi\,\Gamma(1+1/a)}\,\frac{N}{p_\F}\left(1-\frac{p^2}{p_\F^2}\right)^{1/a}
\qquad(\abs{p}<p_\F),
\label{fpower}
\eeq
so that the ballistic late-time profile of a large fermionic cloud reads
\beq
\rho_N(x;t)\approx\frac{\Gamma(3/2+1/a)}{\sqrt\pi\,\Gamma(1+1/a)}\,\frac{N}{p_\F t}\!\left(\!1-\frac{x^2}{(p_\F t)^2}\!\right)^{1/a}
\quad(\abs{x}<p_\F t).
\label{rhopower}
\eeq

Setting $a=2$ in the above results,
we recover the case of a harmonic potential, studied at depth in section~\ref{1dhar}.
We have $p_\F^2=2N\o$, so that~(\ref{fpower}) and~(\ref{rhopower})
coincide with the Wigner semi-circle laws~(\ref{sigmawigner}) and~(\ref{rhowigner}).
Remarkably enough,
for arbitrary values of the exponent $a$ characterising the shape of the potential,
the results~(\ref{fpower}) and~(\ref{rhopower})
can be viewed as deformations of the Wigner semi-circle law,
where the square-root endpoint singularities are replaced by power-law singularities with exponent $1/a$.

The moments of the positions of all fermions, defined in~(\ref{mudef}),
are studied in all space dimensions in section~\ref{hipower}.

\subsection{Square-well potential}
\label{1dsquare}

We now consider the case of a square-well potential:
\beq
V(x)=\left\{\matrix{
0\hfill&(\abs{x}<L/2),\cr
+\infty\quad&(\abs{x}>L/2).
}\right.
\label{1dwell}
\eeq
The particles are thus confined in an interval of length $L$, between two impenetrable walls at $x=\pm L/2$,
imposing Dirichlet boundary conditions.

The square-well potential~(\ref{1dwell}) can be viewed as the $a\to\infty$ limit of the power-law potential~(\ref{vpower}).
The formulas~(\ref{fpower}) and~(\ref{rhopower}) predict that the distributions $f_N(p)$ and $\rho_N(x;t)$
become asymptotically uniform in the intervals $\abs{p}<p_\F$ and $\abs{x}<p_\F t$.
This asymptotic prediction stemming from the semi-classical analysis performed in section~\ref{1dlate}
can be corroborated by exact calculations at finite $N$.
The square-well case is indeed exactly solvable,
in the sense that all wavefunctions are known explicitly.
At variance with the case of a harmonic potential, the many-body dynamics is however not exactly tractable.
The energy levels read
\beq
E_n=(n+1)^2\frac{\pi^2}{2L^2}\qquad(n=0,1,\dots).
\eeq
The corresponding wavefunctions are as follows.

\noindent $\bullet$
Even sector ($n=2k$, $k=0,1,\dots$):
\beqa
\psi_{2k}(x)&=&\sqrt{\frac{2}{L}}\,\cos\frac{(2k+1)\pi x}{L},
\nonumber\\
\h\psi_{2k}(p)&=&(-1)^k\sqrt{\frac{L}{\pi^3}}\,\frac{2(2k+1)}{(2k+1)^2-z^2}\,\cos\frac{\pi z}{2},
\label{phieven}
\eeqa

\noindent $\bullet$
Odd sector ($n=2k-1$, $k=1,2,\dots$):
\beqa
\psi_{2k-1}(x)&=&\sqrt{\frac{2}{L}}\,\sin\frac{2k\pi x}{L},
\nonumber\\
\h\psi_{2k-1}(p)&=&(-1)^k\sqrt{\frac{L}{\pi^3}}\,\frac{4\ii k}{4k^2-z^2}\,\sin\frac{\pi z}{2},
\label{phiodd}
\eeqa
with
\beq
z=\frac{pL}{\pi}.
\eeq

The momentum distribution $f_N(p)$, given by~(\ref{sgal}), can be worked out explicitly for finite $N$
(see~\ref{wellderiv}).
Its expression depends on the parity of $N$, according~to
\beqa
&&f_{2K}(p)=\frac{L}{4\pi^3}\left(\cos^2\frac{\pi z}{2}\,S_\E(K)+\sin^2\frac{\pi z}{2}\,S_\O(K+1)\right),
\label{fnwell}
\\
&&f_{2K+1}(p)=\frac{L}{4\pi^3}\left(\cos^2\frac{\pi z}{2}\,S_\E(K+1)+\sin^2\frac{\pi z}{2}\,S_\O(K+1)\right),
\nonumber
\eeqa
and
\beqa
S_\E(K)
&=&\psi'\!\left(\frac{1+z}{2}\right)-\psi'\!\left(K+\frac{1+z}{2}\right)
\nonumber\\
&+&\psi'\!\left(\frac{1-z}{2}\right)-\psi'\!\left(K+\frac{1-z}{2}\right)
\nonumber\\
&+&\frac{2}{z}\left(\psi\!\left(\frac{1+z}{2}\right)-\psi\!\left(K+\frac{1+z}{2}\right)\right)
\nonumber\\
&-&\frac{2}{z}\left(\psi\!\left(\frac{1-z}{2}\right)-\psi\!\left(K+\frac{1-z}{2}\right)\right),
\label{seres}
\\
S_\O(K)
&=&\psi'\!\left(1+\frac{z}{2}\right)-\psi'\!\left(K+\frac{z}{2}\right)
\nonumber\\
&+&\psi'\!\left(1-\frac{z}{2}\right)-\psi'\!\left(K-\frac{z}{2}\right)
\nonumber\\
&+&\frac{2}{z}\left(\psi\!\left(1+\frac{z}{2}\right)-\psi\!\left(K+\frac{z}{2}\right)\right)
\nonumber\\
&-&\frac{2}{z}\left(\psi\!\left(1-\frac{z}{2}\right)-\psi\!\left(K-\frac{z}{2}\right)\right),
\label{sores}
\eeqa
where $\psi$ and $\psi'$ are the digamma and trigamma functions.

\begin{figure}[!htbp]
\begin{center}
\includegraphics[angle=0,width=.65\linewidth,clip=true]{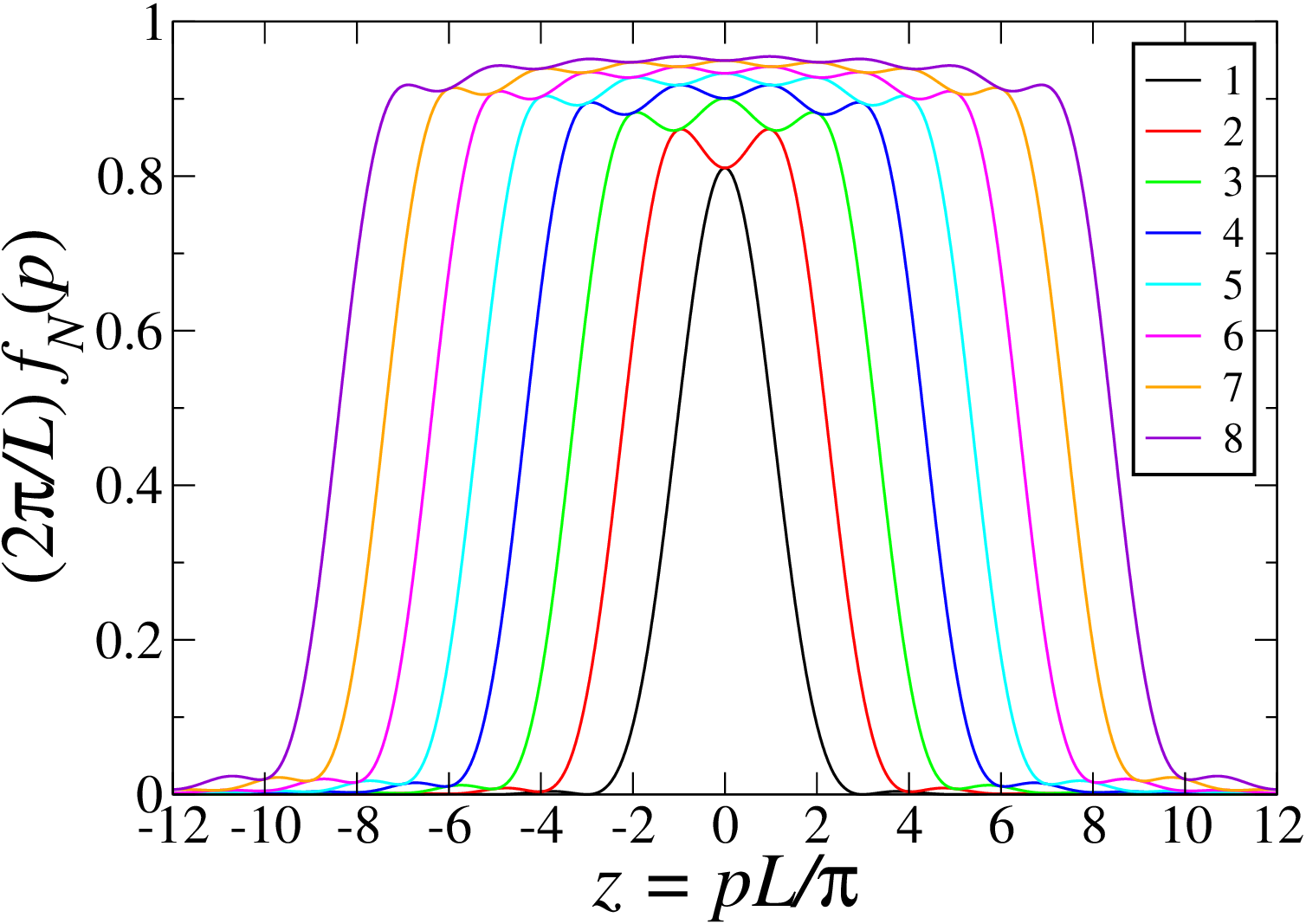}
\caption{
Rescaled momentum distribution $(2\pi/L)f_N(p)$ for the one-dimensional square well,
plotted against $z=pL/\pi$ for $N=1$ to 8 (see legend).}
\label{fsquare}
\end{center}
\end{figure}

When $K$ is infinitely large,
using the identities
\beqa
&&\psi(z)-\psi(1-z)=-\pi\cot\pi z,
\nonumber\\
&&\psi'(z)+\psi'(1-z)=\frac{\pi^2}{\sin^2\pi z}
\eeqa
stemming from the reflection formula for the Gamma function,
\beq
\Gamma(z)\Gamma(1-z)=\frac{\pi}{\sin\pi z},
\eeq
we obtain
\beqa
S_\E(\infty)&=&\frac{\pi^2}{\cos^2\frad{\pi z}{2}}+\frac{2\pi}{z}\tan\frac{\pi z}{2},
\nonumber\\
S_\O(\infty)&=&\frac{\pi^2}{\sin^2\frad{\pi z}{2}}-\frac{2\pi}{z}\cot\frac{\pi z}{2},
\eeqa
so that the expressions~(\ref{fnwell}) boil down to
\beq
f_N(p)\to\frac{L}{2\pi}\qquad(\abs{p}<p_\F),
\label{fasywell}
\eeq
and zero elsewhere,
with
\beq
p_\F\approx\frac{N\pi}{L}.
\eeq
The uniform momentum distribution~(\ref{fasywell}) agrees with the semi-classical analysis performed in section~\ref{1dlate}.
Figure~\ref{fsquare} shows how successive individual levels contribute to build up the latter uniform distribution.

\begin{figure}[!htbp]
\begin{center}
\includegraphics[angle=0,width=.65\linewidth,clip=true]{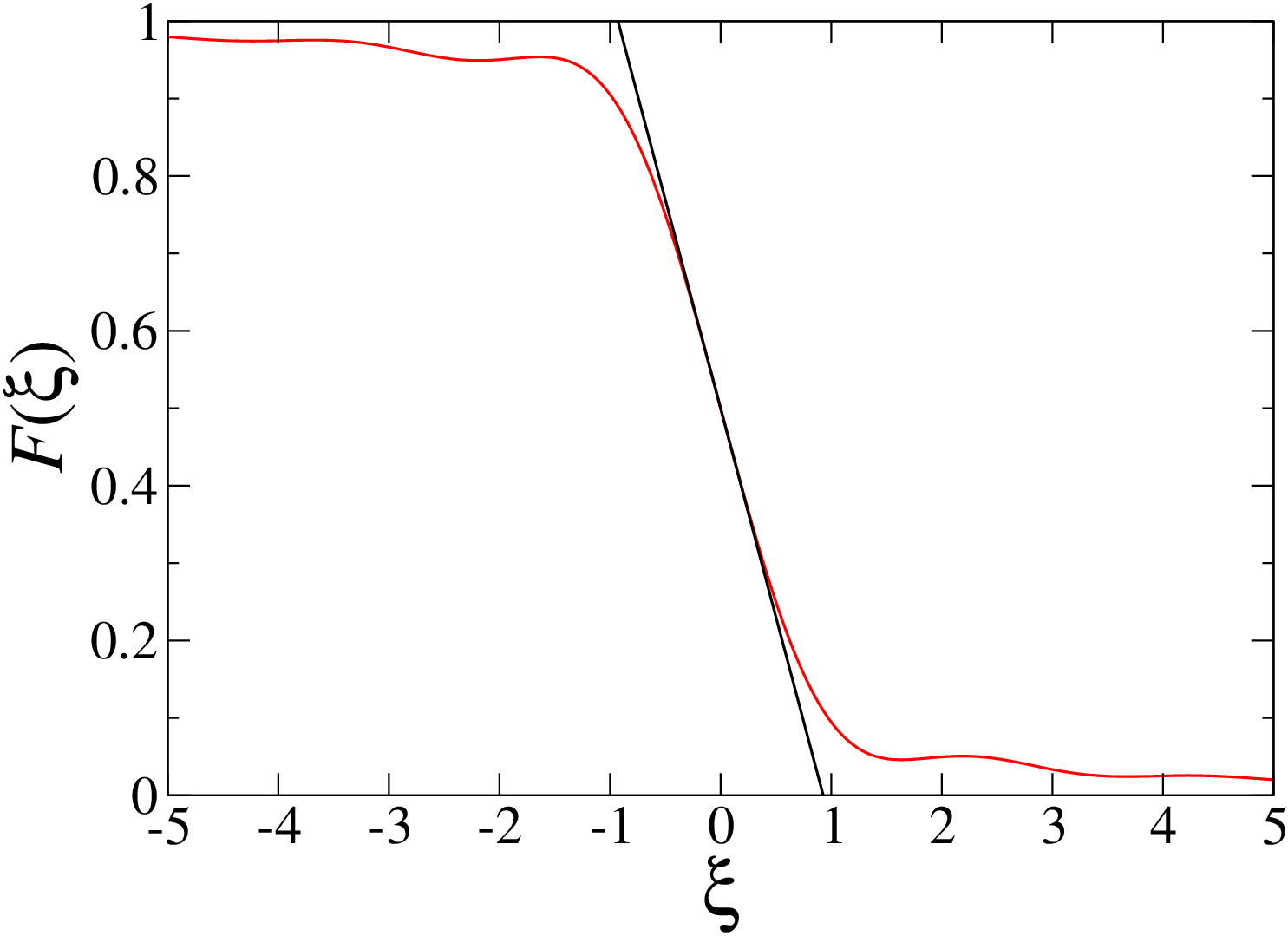}
\caption{
Scaling function $F(\xi)$ entering the formula~(\ref{scawell}) describing the drop in the momentum distribution near $p_\F$.
Black line: inflection tangent at the origin~(see~(\ref{Fzero})).}
\label{Fplot}
\end{center}
\end{figure}

In the vicinity of $p_\F$, the momentum distribution $f_N(p)$ drops from the constant value~(\ref{fasywell}) to zero
over a microscopic range of values of $p$.
Setting
\beq
z=N+\half+\xi,\qquad\hbox{i.e.,}\quad p=p_\F+\frac{\pi}{L}\left(\half+\xi\right),
\eeq
both formulas~(\ref{fnwell}) yield after some algebra
\beq
f_N(p)\approx\frac{L}{2\pi}\;F(\xi),
\label{scawell}
\eeq
with
\beq
F(\xi)=\half
+\frac{1+\sin\pi\xi}{4\pi^2}\,\psi'\!\left(\frac{1}{4}+\frac{\xi}{2}\right)
-\frac{1-\sin\pi\xi}{4\pi^2}\,\psi'\!\left(\frac{1}{4}-\frac{\xi}{2}\right).
\label{Fdef}
\eeq
The scaling function $F(\xi)$ interpolates between $F(-\infty)=1$ and $F(+\infty)=0$.
It obeys $F(-\xi)=1-F(\xi)$, so that $F(\xi)-1/2$ is an odd function of $\xi$.
Figure~\ref{Fplot} shows that most of its decay is nearly linear, i.e., $F(\xi)\approx F(0)+F'(0)\xi$, with
\beq
F(0)=\half,\qquad F'(0)=\frac{\psi'(1/4)}{2\pi}+\frac{\psi''(1/4)}{4\pi^2}\approx-0.538869.
\label{Fzero}
\eeq
The decay of $F(\xi)$ at large positive $\xi$, namely
\beq
F(\xi)=\frac{1}{\pi^2}\left[\frac{1}{\xi}-\frac{1}{12\xi^3}+\cdots
+\sin\pi\xi\left(\frac{1}{2\xi^2}-\frac{3}{8\xi^4}+\cdots\right)\right],
\eeq
is slow and oscillating.
The leading behavior $F(\xi)\approx1/(\pi^2\xi)$ causes
the mean number of fermions with momenta outside the interval $\abs{p}<p_\F$ to grow logarithmically~as
\beq
N_{\rm out}\approx\frac{\ln N}{\pi^2}.
\label{Nlog}
\eeq

\section{The higher-dimensional case}
\label{hid}

\subsection{General setting}
\label{hidgal}

We now consider $N$ non-interacting fermions in arbitrary higher dimension ($d\ge2$),
with spin degeneracy $g$,
subjected to an isotropic (i.e., radial) scalar potential.
The one-body Hamiltonian reads
\beq
\H=\frac{\vp^2}{2}+V(r),
\eeq
with $\vp=-\ii\Nabla$ and $r=\abs{\vx}$.
The potential $V(r)$ is again assumed to grow at infinity,
so that the spectrum of $\H$ is entirely discrete.
Besides spin degeneracy, the eigenstates of $\H$ also have an orbital degeneracy.

As mentioned above, one-dimensional fermions can be thought of as spinless.
In other physically relevant dimensions ($d=2$ or 3), fundamental fermions have spin $S=1/2$ and spin degeneracy $g=2$.
In dimension $d=2$, an eigenstate with energy $E$ and angular momentum $l\ge0$
reads $\psi_{lj}(\vx)=\e^{\pm\ii l\theta}\psi_{lj}(r)$,
where the radial wavefunction $\psi_{lj}(r)$ obeys
\beq
-\half\left(\psi_{lj}''(r)+\frac{1}{r}\,\psi_{lj}'(r)\right)+\left(\frac{l^2}{r^2}+V(r)\right)\psi_{lj}(r)=E\psi_{lj}(r).
\eeq
Its orbital degeneracy reads $g(0)=1$ and $g(l)=2$ for $l>0$.
In dimension $d=3$, an eigenstate with energy $E$ and angular quantum numbers $(l,m)$,
with $l\ge0$ and $\abs{m}\le l$,
reads $\psi_{lmj}(\vx)=Y_{lm}(\theta,\varphi)\psi_{lj}(r)$,
where the radial wavefunction $\psi_{lj}(r)$ obeys
\beq
-\half\left(\psi_{lj}''(r)+\frac{2}{r}\,\psi_{lj}'(r)\right)+\left(\frac{l(l+1)}{r^2}+V(r)\right)\psi_{lj}(r)=E\psi_{lj}(r).
\eeq
Its orbital degeneracy reads $g(l)=2l+1$ for all $l\ge0$.

We again assume that the system is prepared in one of its many-body ground states, so that
the wavefunction is again given by a Slater determinant, namely
\beqa
\braket{\Psi(0)}{\vxx}&=&\frac{1}{\sqrt{N!}}\,\det(\psi_{k-1}(\vx_l)),
\nonumber\\
\braket{\Psi(0)}{\vpp}&=&\frac{1}{\sqrt{N!}}\,\det(\h\psi_{k-1}(\vp_l)),
\eeqa
where the index $k$ gathers spin and orbital quantum numbers.
The main novelty with respect to the one-dimensional situation is the generic occurrence of degeneracies.
As a consequence, there is no canonical way of defining a basis of eigenstates $\psi_k$,
and therefore no unique many-body ground state for a generic fermion number $N$.
In other words, $\ket{\Psi(0)}$ depends on details of how the system is prepared.
It however becomes unique and isotropic whenever the initial state consists of complete shells.
This can only occur for some specific fermion numbers $N$.

\subsection{Late-time regime}
\label{hilate}

Let us again focus our attention onto the late-time regime of the expansion dynamics.
The line of reasoning put forward in section~\ref{1dlate} in the one-dimensional situation still
holds in the higher-dimensional setting.
The fermions expand ballistically according~to
\beq
\vx_n\approx\vp_n\,t.
\label{hibal}
\eeq
Furthermore, interferences again become negligible,
so that the momenta $\vp_n$ can be viewed as independent classical random variables, whose distribution
\beq
f_N(\vp)=\sum_{n=0}^{N-1}\abs{\h\psi_n(\vp)}^2,
\label{higal}
\eeq
normalized as
\beq
\int f_N(\vp)\,\dd^d\vp=N,
\label{hifnor}
\eeq
inherits the non-universal features of the initial state $\ket{\Psi(0)}$.

When the fermion number $N$ is very large,
most of the one-body states involved in~(\ref{higal}) are highly excited.
The ensuing distribution $f_N(\vp)$ can again be evaluated by semi-classical methods.
In the present higher-dimensional setting, in arbitrary space dimension $d$, this goes as follows.

The semi-classical estimate for the integrated density of states~$n(E)$,
i.e., the number of quantum states below some energy $E$, counted with their multiplicities,
is often referred to as the Thomas-Fermi formula (see e.g.~\cite{Berry,Eckhardt,LL}).
This reads
\beq
n(E)\approx gW(E),
\label{hine}
\eeq
where $W(E)$ is the phase-space volume below energy $E$,
in units of $h^d$, i.e., $(2\pi)^d$,
namely\footnote{$\1()$ denotes the characteristic function of a set
(here, defined by an inequality), equal to unity in the set and to zero else.}
\beqa
W(E)
&=&\int\!\!\!\int\frac{\dd^d\vx\,\dd^d\vp}{(2\pi)^d}\ \1\!\left(\frac{\vp^2}{2}+V(\vx)<E\right)
\nonumber\\
&=&\frac{S_d^2}{(2\pi)^d}\int_0^\infty r^{d-1}\dd r\int_0^\infty p^{d-1}\dd p\ \1\!\left(\frac{p^2}{2}+V(r)<E\right)
\nonumber\\
&=&\frac{S_d^2}{d(2\pi)^d}\int_0^\infty r^{d-1}\dd r\,\left(2(E-V(r))\right)^{d/2},
\label{wres}
\eeqa
where
\beq
S_d=\frac{2\pi^{d/2}}{\Gamma(d/2)}
\eeq
is the area of the unit sphere in dimension $d$.

The classical microcanonical measure at energy $E$ reads
\beq
f(\vx,\vp;E)=\frac{1}{(2\pi)^d\,Z(E)}\ \delta\!\left(\frac{\vp^2}{2}+V(\vx)-E\right),
\label{himic}
\eeq
where the microcanonical partition function is nothing but
\beq
Z(E)=W'(E).
\label{hiz}
\eeq
The expression~(\ref{himic}) is isotropic in both $\vx$ and $\vp$ spaces.
We have therefore
\beqa
f(\vp;E)
&=&\frac{1}{Z(E)}\int\frac{\dd^d\vx}{(2\pi)^d}\ \delta\!\left(\frac{p^2}{2}+V(\vx)-E\right)
\nonumber\\
&=&\frac{S_d}{(2\pi)^d\,Z(E)}\int_0^\infty r^{d-1}\dd r\ \delta\!\left(\frac{p^2}{2}+V(r)-E\right)
\nonumber\\
&=&\frac{S_d}{(2\pi)^d\,Z(E)}\,\frac{r_0^{d-1}}{V'(r_0)},
\label{hifpe}
\eeqa
with $p^2=2(E-V(r_0))$.

The momentum distribution $f_N(\vp)$ introduced in~(\ref{higal}) can now be estimated as
\beq
f_N(\vp)\approx\int_0^Nf(\vp;E)\,\dd n,
\eeq
where $n$ and $E$ are related by~(\ref{hine}).
This relation, together with~(\ref{hiz}), implies $\dd n\approx g\,Z(E)\,\dd E$.
Introducing the Fermi momentum $p_\F$ and energy $E_\F=p_\F^2/2$ such that
\beq
N=n(E_\F)\approx gW(E_\F),
\label{nome}
\eeq
and using~(\ref{hifpe}),
we are left with the following expression
\beq
f_N(\vp)\approx\frac{gS_d}{(2\pi)^d}\int_{p^2/2}^{E_\F}\frac{r_0^{d-1}}{V'(r_0)}\,\dd E\qquad(p=\abs{\vp}<p_\F),
\label{hisemi}
\eeq
where $r_0$ depends on $p$ and $E$ through $p^2=2(E-V(r_0))$.
The expression~(\ref{hisemi}) is a generalization of~(\ref{fnsemi}) in arbitrary dimension $d$.
It is isotropic, so that~(\ref{hifnor}) reads
\beq
S_d\int_0^{p_\F}p^{d-1} f_N(p)\,\dd p=N.
\label{hisonor}
\eeq
As a consequence of the ballistic scaling~(\ref{hibal}),
the result~(\ref{hisemi}) predicts that the profile of a large fermionic cloud,
\beq
\rho_N(\vx;t)\approx\frac{1}{t^d}\,f_N\!\left(\frac{r}{t}\right),
\eeq
isotropizes at late times,
and is essentially supported by a ball whose radius expands ballistically as $R(t)\approx p_\F t$.

\subsection{Power-law potential}
\label{hipower}

Let us give the explicit forms of the above general results
in the case of a radial power-law potential of the form
\beq
V(r)=\left(\frac{r}{\ell}\right)^a,
\eeq
where $a>0$ is an arbitrary exponent, and $\ell$ again measures the trap size.

The expression~(\ref{wres}) for the phase space volume below energy $E$ evaluates to
\beq
W(E)=\frac{\Gamma(1+d/a)}{2^{d/2}\Gamma(1+d/a+d/2)\Gamma(1+d/2)}\,\ell^d\,E^{d/a+d/2},
\eeq
so that~(\ref{nome}) yields
\beq
E_\F=\frac{p_\F^2}{2}\approx
\left(\frac{2^{d/2}\Gamma(1+d/a+d/2)\Gamma(1+d/2)}{\Gamma(1+d/a)}\,\frac{N}{g\ell^d}\right)^{\frat{2a}{(a+2)d}}.
\label{hief}
\eeq
The estimate~(\ref{hisemi}) for the momentum distribution evaluates to
\beq
f_N(p)\approx\frac{\Gamma(1+d/a+d/2)}{\pi^{d/2}\,\Gamma(1+d/a)}\,\frac{N}{p_\F^d}\left(1-\frac{p^2}{p_\F^2}\right)^{d/a}
\qquad(\abs{p}<p_\F),
\label{hifpower}
\eeq
so that the ballistic late-time profile of a large fermionic cloud reads
\beq
\rho_N(r;t)\approx\frac{\Gamma(1+d/a+d/2)}{\pi^{d/2}\,\Gamma(1+d/a)}\,\frac{N}{(p_\F t)^d}\left(1-\frac{r^2}{(p_\F t)^2}\right)^{d/a}.
\label{hirhopower}
\eeq

The results~(\ref{hief}),~(\ref{hifpower}) and~(\ref{hirhopower})
generalize~(\ref{1def}),~(\ref{fpower}) and~(\ref{rhopower}) to arbitrary dimension $d$ and spin degeneracy $g$.
Remarkably enough,
in any dimension $d$ and for arbitrary values of the exponent $a$,
the expressions~(\ref{hifpower}) and~(\ref{hirhopower})
again appear as deformations of the Wigner semi-circle law,
where the square-root endpoint singularities are replaced by power-law singularities with exponent $d/a$.
In the $a\to\infty$ limit, the problem at stake becomes that of an infinite spherical well.
Particles execute free motion inside a spherical billiard $(r<R)$ with Dirichlet boundary condition.
The formula~(\ref{hifpower}) predicts that the radial distribution $f_N(p)$
is asymptotically uniform inside the Fermi ball $p<p_\F$.
This prediction is corroborated by numerical data
for an infinite circular well, to be presented in section~\ref{hidisk}.

The radial moments of the positions of all fermions are defined in analogy with~(\ref{mudef}), namely
\beq
\mu_{2k}(t)
=\sum_{n=1}^N\braopket{\Psi(t)}{\abs{\vx_n}^{2k}}{\Psi(t)}
=S_d\int_0^\infty r^{2k+d-1}\rho_N(r;t)\,\dd r.
\label{himudef}
\eeq
The estimate~(\ref{hirhopower}) implies that these moments grow as
\beq
\mu_{2k}(t)\approx\frac{\Gamma(k+d/2)\Gamma(d/a+d/2+1)}{\Gamma(d/2)\Gamma(k+d/a+d/2+1)}\,N(p_\F t)^{2k},
\label{himuas}
\eeq
i.e.,
\beqa
&&\mu_2(t)\approx\frac{ad}{(a+2)d+2a}\,N(p_\F t)^2,
\nonumber\\
&&\mu_4(t)\approx\frac{a^2d(d+2)}{((a+2)d+2a)((a+2)d+4a)}\,N(p_\F t)^4,
\eeqa
and so on.
When the dimension $d$ gets very large,
the expression~(\ref{himuas}) boils down to
\beq
\mu_{2k}(t)\approx N\left(\frac{a}{a+2}\right)^k(p_\F t)^{2k}.
\eeq
The radial momentum distribution, proportional to the product $p^{d-1}f_N(p)$,
indeed becomes strongly peaked around a spherical shell with radius
\beq
p^*=\sqrt\frac{a}{a+2}\ p_\F.
\eeq

\subsection{The example of an infinite circular well}
\label{hidisk}

This section is devoted to the two-dimensional example of an infinite circular well,
defined by the potential
\beq
V(r)=\left\{\matrix{
0\hfill&(r<1),\cr
+\infty\quad&(r>1).
}\right.
\eeq
Particles execute free motion inside a circular billiard,
whose radius is chosen to be unity for simplicity.
The impenetrable wall imposes a Dirichlet boundary condition at $r=1$.
The wavefunctions of the one-body eigenstates read
\beq
\psi_{lj}(r,\theta)=C_{lj}\,\e^{\pm\ii l\theta}J_l(k_{lj}r),
\eeq
where $J_l(z)$ is the Bessel function of the first kind,
and the momentum $k_{lj}$ is the $j$th positive zero of the latter function, with $l\ge0$ and $j\ge1$.
The above eigenstates have energy
\beq
E_{lj}=\half k_{lj}^2,
\eeq
spin degeneracy $g=2$, and orbital degeneracy $g(0)=1$ and $g(l)=2$ for $l\ge1$.

The ten smallest moments $k_{lj}$ are listed in table~\ref{klj}, together with their multiplicities $2g(l)$.
The fermion numbers corresponding to complete shells are therefore
\beq
N=2,\ 6,\ 10,\ 12,\ 16,\ 20,\ 24,\ 28,\ 30,\ 34,\ \dots
\label{nlist}
\eeq
This list can be viewed as some two-dimensional analogue of the well-known magic numbers of nuclear physics.
One should however bear in mind that the theory of those nuclear magic numbers is more intricate,
involving in particular a strong spin-orbit coupling~\cite{MJ}.

\begin{table}[!htbp]
\begin{center}
$$
\begin{array}{|c|c|c|c|}
\hline
l & j & 2g(l) & k_{lj} \\
\hline
0 & 1 & 2 & 2.404825557 \\
1 & 1 & 4 & 3.831705970 \\
2 & 1 & 4 & 5.135622301 \\
0 & 2 & 2 & 5.520078110 \\
3 & 1 & 4 & 6.380161895 \\
1 & 2 & 4 & 7.015586669 \\
4 & 1 & 4 & 7.588342434 \\
2 & 2 & 4 & 8.417244140 \\
0 & 3 & 2 & 8.653727912 \\
5 & 1 & 4 & 8.771483815 \\
\hline
\end{array}
$$
\caption{First ten moments $k_{lj}$ entering the sum~(\ref{fndisk}), with labels $l$ and $j$,
and multiplicities $2g(l)$.}
\label{klj}
\end{center}
\end{table}

The identity (see~\cite[Eq.~6.521]{GR})
\beq
\int_0^1 J_l(ar)J_l(br)\,r\,\dd r=\frac{a J_l'(a) J_l(b)-b J_l(a)J_l'(b)}{b^2-a^2}
\eeq
enables us to evaluate the normalizing constants
\beq
C_{lj}=\frac{1}{\sqrt{\pi}\,J_l'(k_{lj})}
\eeq
and the wavefunctions in Fourier space, namely
\beq
\h\psi_{lj}(\vp)=\frac{k_{lj}}{p^2-k_{lj}^2}\,\frac{J_l(p)}{\sqrt\pi},
\eeq
for $\vp$ along the positive $x$-axis, with $p=\abs{\vp}$.

For fermion numbers $N$ corresponding to complete shells (see~(\ref{nlist})),
the expression~(\ref{higal}) of the momentum distribution takes the isotropic form
\beq
f_N(p)=\frac{1}{\pi}\sum_{lj}\biggl(\frac{k_{lj}}{p^2-k_{lj}^2}\biggr)^2J_l^2(p),
\label{fndisk}
\eeq
where each eigenstate is repeated according to its multiplicity $2g(l)$, namely 2 for $l=0$ and 4 for $l\ge1$.

At zero momentum, the sum entering~(\ref{fndisk}) only involves the zeros $k_{0j}$ of~$J_0$.
The sum rule (see e.g.~\cite{watson})
\beq
\sum_{j\ge1}\frac{1}{k_{0j}^2}=\frac{1}{4}
\eeq
yields the limit
\beq
f_N(0)\to\frac{1}{2\pi}.
\label{fncirc0}
\eeq

\begin{figure}[!htbp]
\begin{center}
\includegraphics[angle=0,width=.65\linewidth,clip=true]{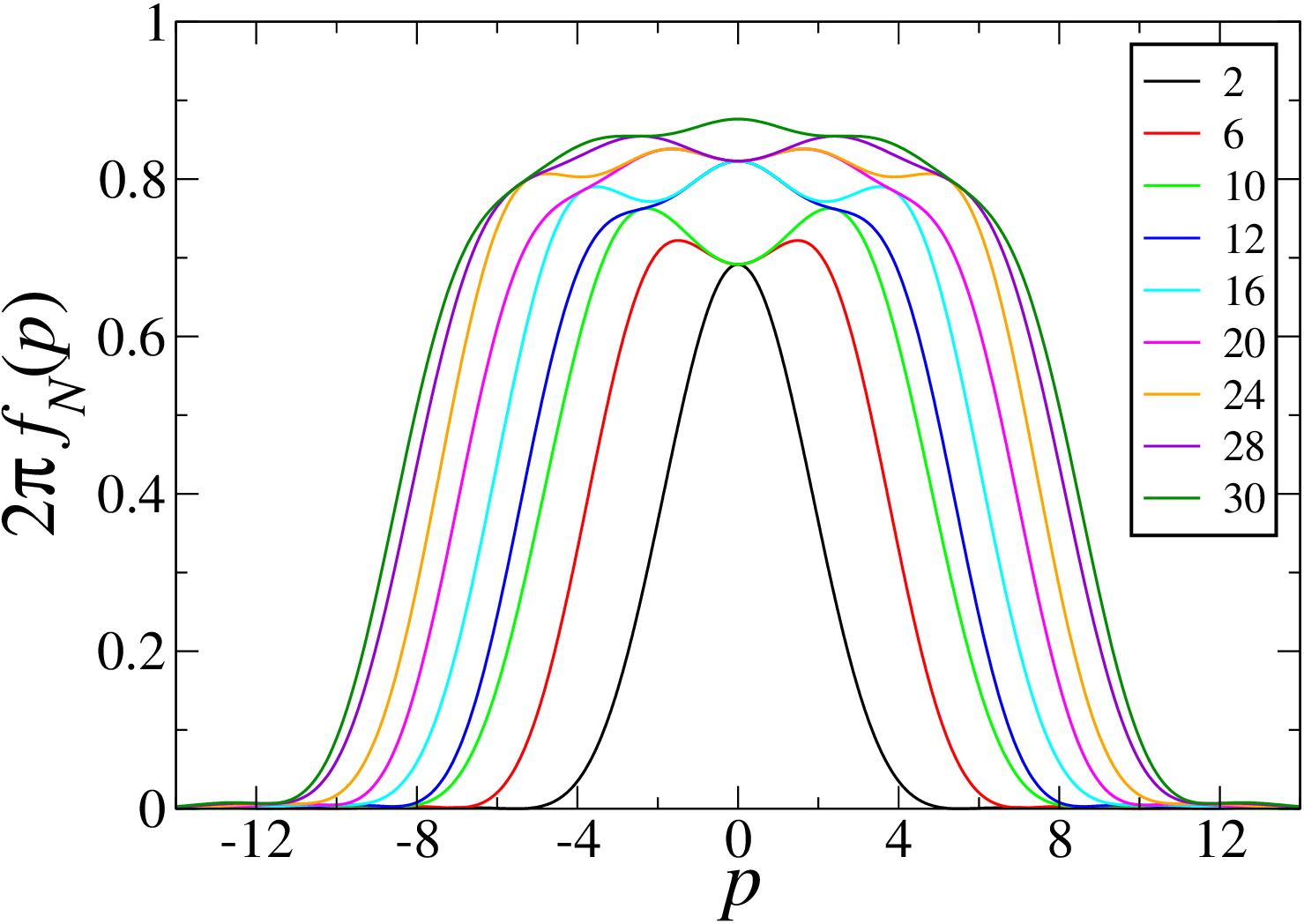}
\caption{
Rescaled momentum distribution $2\pi\,f_N(p)$ for the circular billiard,
plotted against $p$ for the lowest nine values of $N$ corresponding to complete shells (see~(\ref{nlist})).}
\label{fdisk}
\end{center}
\end{figure}

To leading order at large $N$, the Fermi momentum $p_\F$ can be derived
by means of the semi-classical analysis of section~\ref{hilate}.
The expression~(\ref{wres}) of the phase space volume yields $W(E)=E/2$, and so
\beq
p_\F^2\approx 2N.
\label{pfcirc}
\eeq
As a consequence of the above, we have $\pi p_\F^2 f_N(0)\approx N$,
in agreement with the prediction
that the momentum distribution $f_N(p)$ is asymptotically uniformly equal to $1/(2\pi)$
inside the Fermi disk $p<p_\F$, and obeys the normalization~(\ref{hisonor}).
Figure~\ref{fdisk} shows how successive shells contribute to build up the latter uniform distribution.

A general feature of higher-dimensional quantum systems
is that the number $n(E)$ of eigenstates below some energy $E$
can be written as the sum of a smooth component and an oscillatory one
(see e.g.~\cite{Berry,Eckhardt}, and~\cite{BH} for an overview):
\beq
n(E)=\overline n(E)+n_\osc(E).
\eeq
For a regular enough two-dimensional domain with area $A$ and perimeter length $P$,
the smooth component has the following expansion
\beq
\overline n(E)=\frac{AE}{\pi}\pm\frac{P\sqrt{E}}{\pi\sqrt2}+\cdots,
\eeq
where the upper (resp.~lower) sign refers to Neumann (resp.~Dirichlet) boundary conditions,
and the first unwritten term is a constant `curvature' term.
In the present setting, this yields
\beq
\overline n(E)\approx E-\sqrt{2E}.
\eeq
The oscillatory component $n_\osc(E)$ of the integrated density of states
can be expressed in terms of the classical periodic orbits~\cite{Balian,G1,G2,B1,B2}.
Its dependence on energy is usually quite irregular.

In order to compare the above theoretical predictions to actual data,
we introduce two different operational definitions of the Fermi momentum:

\begin{enumerate}

\item[(1)]
The Fermi momentum denoted by $p_{\F 1}$
is defined as being the largest of all momenta $k_{lj}$ entering the sum~(\ref{fndisk}) for any given $N$.
Neglecting oscillations, we have the estimate $N\approx\overline n(E_{\F 1})$, with $E_{\F 1}=p_{\F 1}^2/2$,
and so
\beq
p_{\F 1}\approx\sqrt{2N}+1.
\label{pfun}
\eeq
The leading term agrees with~(\ref{pfcirc}), as should be.

\item[(2)]
The Fermi momentum denoted by $p_{\F 2}$ is defined, for $N$ corresponding to complete shells (see~(\ref{nlist})),
as being the value of $p$ at which the distribution~$f_N(p)$ is in the middle of its fall-off
from its plateau value~(\ref{fncirc0}) to zero, i.e., $f_N(p_{\F 2})=1/(4\pi)$.

\end{enumerate}

\begin{figure}[!htbp]
\begin{center}
\includegraphics[angle=0,width=.65\linewidth,clip=true]{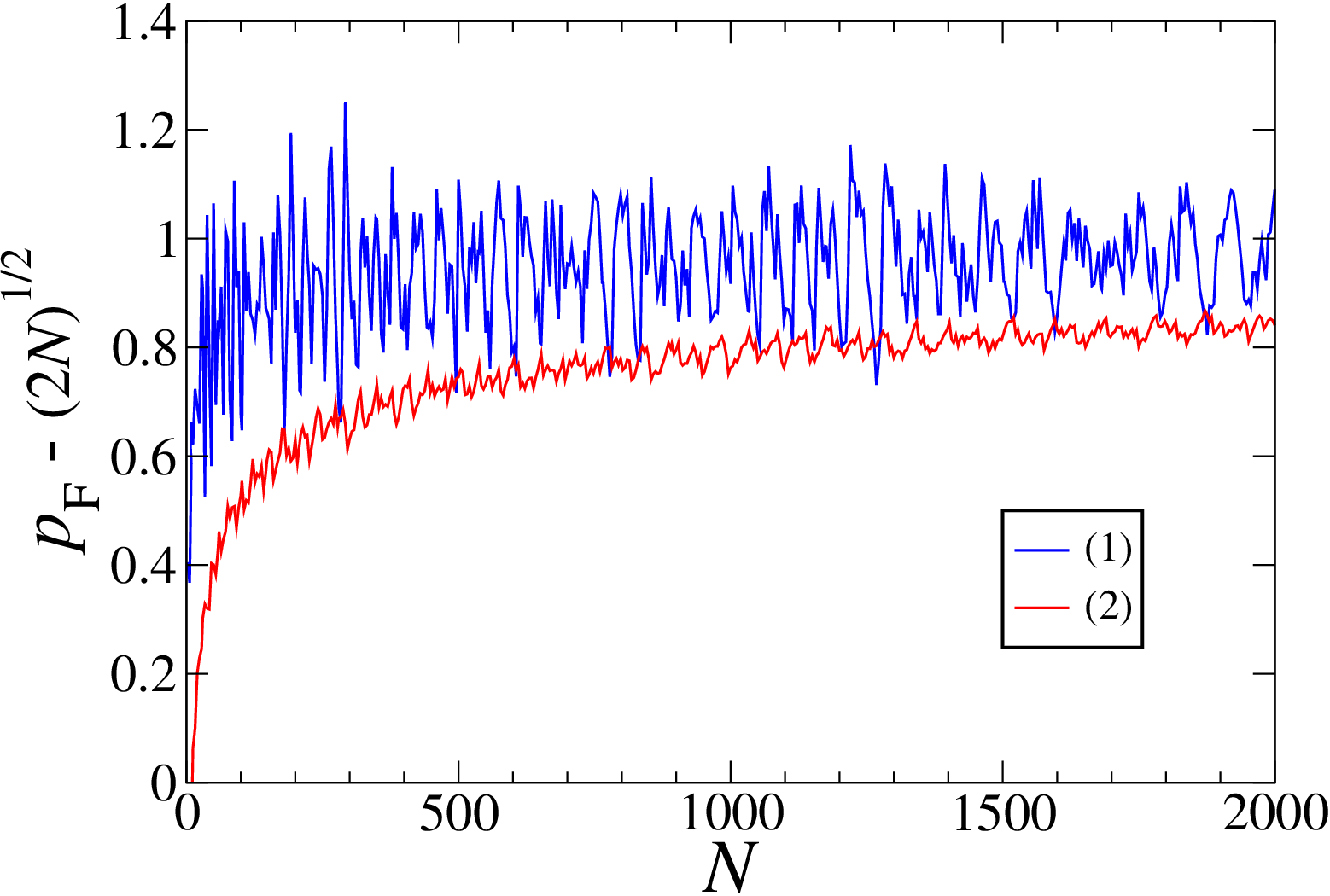}
\caption{
Differences $p_{\F 1}-\sqrt{2N}$ (blue) and $p_{\F 2}-\sqrt{2N}$ (red),
plotted against all fermion numbers $N$ listed in~(\ref{nlist}) up to 2000.}
\label{pfs}
\end{center}
\end{figure}

Figure~\ref{pfs} shows plots of the differences $p_{\F 1}-\sqrt{2N}$ (blue) and $p_{\F 2}-\sqrt{2N}$ (red)
against all fermion numbers~$N$ corresponding to complete shells (see~(\ref{nlist})) up to 2000.
The first difference approaches unity, as predicted by~(\ref{pfun}),
end exhibits rather large oscillations.
The second difference (red) likely also converges to a limit,
exhibiting both a stronger finite-size correction and smaller oscillations.

\section{Discussion}
\label{disc}

The main findings of this work have been summarized in the Introduction.
In a nutshell, the late-time ballistic expansion of a large cloud of non-interacting fermions after a sudden release
is investigated by means of semi-classical methods.
The main focus is on isotropic power-law trapping potentials with arbitrary exponent~$a$ in any space dimension $d$.
The initial momentum distribution is confined to the Fermi ball $(p=\abs{\vp}<p_\F)$,
and so the fermionic cloud is essentially confined to a ball whose radius expands ballistically $(r=\abs{\vx}<p_\F t)$.
The expressions~(\ref{hifpower}) of the momentum distribution
and~(\ref{hirhopower}) of the asymptotic profile of the fermionic cloud exhibit a universal edge exponent $d/a$,
thus generalizing the Wigner semi-circle law and the Thomas-Fermi distribution,
which are respectively recovered for $(d,a)=(1,2)$ and~(3,2).

Besides the present setting of free fermions,
expanding profiles with bounded supports and singularities at moving edges
are met in a variety of classical and quantum-mechanical situations
(see~\cite{Barenblatt} for an overview).
The expansion of over\-damped stochastic Riesz gases of classical particles subjected to long-range power-law interactions
has been studied lately by means of a non-local hydrodynamic formalism
in various dimensions~\cite{KM,FRH}.
There, density profiles spread either sub-diffusively or super-diffusively,
and have bounded supports and power-law edge singularities.
In the quantum realm, a single tight-binding particle launched from the origin of a one-dimensional lattice
spreads over a ballistically growing interval, and manifests sharp ballistic fronts near the moving edges of that interval,
whose width grows as $t^{1/3}$~\cite{toro}.
A similar but richer phenomenology,
with sharp fronts both in the bulk and near the endpoints of the spreading zone,
has been reported for quantum walks executed
by bound states of two bosonic or fermionic particles in one dimension~\cite{KLMwalks}.

This work has also unveiled some more detailed specific features.
In the one-dimensional situation,
the momentum distribution~$f_N(p)$ in the vicinity of the Fermi momentum~$p_\F$
manifests a variety of possible behaviors.
In the case of a harmonic trap,
the edge of the Wigner semi-circle law is broadened
into a smooth crossover whose relative extent shrinks as $N^{-2/3}$,
where the momentum distribution obeys the scaling law~(\ref{sigmaairy}),
involving the non-trivial scaling function $A(\xi)$, given by~(\ref{Adef}) and plotted in figure~\ref{Aplot}.
The mean number of fermions outside the Wigner interval,
$N_{\rm out}\approx0.061258$, is a tiny constant.
For the square-well potential,
the momentum distribution drops from a constant bulk value to zero over a microscopic range of $p$.
This drop is described by the non-trivial scaling function $F(\xi)$,
given by~(\ref{Fdef}) and plotted in figure~\ref{Fplot}.
As a consequence of the sharpness of the potential,
$F(\xi)$ has a long tail falling off as $1/\xi$,
so that $N_{\rm out}$ grows logarithmically with the fermion number~$N$, according to~(\ref{Nlog}).
Evaluating the mean number of fermions outside the moving edge of the expanding profile
for more general trapping potentials and in higher dimension remains a challenging open question.

The two-dimensional example of an infinite circular well
exhibits another ubiquitous feature of higher-dimensional bound states,
namely the occurrence of an irregular oscillatory component in the integrated density of states,
and the ensuing irregular dependence of the Fermi momentum $p_\F$ on the fermion number $N$,
even if the latter is restricted to complete shells of degenerate one-body states.
This point is illustrated in figure~\ref{pfs},
showing plots of two different microscopic definitions of the Fermi momentum against $N$,
defined in terms of the largest one-body energy and of the fall-off of the momentum distribution.

\section*{Data availability statement}

Data sharing not applicable to this article.

\section*{Conflict of interest}

The authors declare no conflict of interest.

\appendix

\section{Derivation of the formulas~(\ref{fnwell})}
\label{wellderiv}

This appendix is devoted to the derivation of the expressions~(\ref{fnwell})
of the momentum distribution $f_N(p)$, defined by~(\ref{sgal}), namely
\beq
f_N(p)=\sum_{n=0}^{N-1}\abs{\h\psi_n(p)}^2.
\label{sappgal}
\eeq
We consider for definiteness an even value $N=2K$ of the fermion number.
The formulas~(\ref{phieven}),~(\ref{phiodd}) suggest
to split the contributions of even and odd values of $n$ according to
\beq
f_{2K}(p)=f_{2K,\E}(p)+f_{2K,\O}(p).
\eeq
Setting
\beq
z=\frac{pL}{\pi},
\eeq
we have
\beqa
f_{2K,\E}(p)
&=&\sum_{k=0}^{K-1}\abs{\h\psi_{2k}(p)}^2
\nonumber\\
&=&\frac{L}{\pi^3}\cos^2\frac{\pi z}{2}\sum_{k=0}^{K-1}\left(\frac{2(2k+1)}{(2k+1)^2-z^2}\right)^2.
\eeqa
Using the identity
\beq
\frac{2(2k+1)}{(2k+1)^2-z^2}=\frac{1}{2k+1+z}+\frac{1}{2k+1-z},
\eeq
we have
\beq
f_{2K,\E}(p)=\frac{L}{4\pi^3}\cos^2\frac{\pi z}{2}\,S_\E(K),
\eeq
with
\beqa
S_\E(K)
&=&4\sum_{k=0}^{K-1}\left(\frac{1}{2k+1+z}+\frac{1}{2k+1-z}\right)^2
\nonumber\\
&=&4\sum_{k=0}^{K-1}\left(\frac{1}{(2k+1+z)^2}+\frac{1}{(2k+1-z)^2}\right)
\nonumber\\
&+&8\sum_{k=0}^{K-1}\frac{1}{(2k+1+z)(2k+1-z)}.
\eeqa
Using the identities
\beqa
&&\sum_{k=0}^{K-1}\frac{1}{k+z}=\psi(K+z)-\psi(z),
\nonumber\\
&&\sum_{k=0}^{K-1}\frac{1}{(k+z)^2}=\psi'(z)-\psi'(K+z),
\eeqa
we obtain
\beqa
S_\E(K)
&=&\psi'\!\left(\frac{1+z}{2}\right)-\psi'\!\left(K+\frac{1+z}{2}\right)
\nonumber\\
&+&\psi'\!\left(\frac{1-z}{2}\right)-\psi'\!\left(K+\frac{1-z}{2}\right)
\nonumber\\
&+&\frac{2}{z}\left(\psi\!\left(\frac{1+z}{2}\right)-\psi\!\left(K+\frac{1+z}{2}\right)\right)
\nonumber\\
&-&\frac{2}{z}\left(\psi\!\left(\frac{1-z}{2}\right)-\psi\!\left(K+\frac{1-z}{2}\right)\right),
\eeqa
which is~(\ref{seres}).
Dealing similarly with odd values of $N$ and/or $n$, we obtain~(\ref{fnwell}).

\section*{Orcid ids}

\noindent Pavel Krapivsky: https://orcid.org/0000-0003-3470-5095

\noindent Jean-Marc Luck: https://orcid.org/0000-0003-2151-5057

\section*{References}

\bibliography{paper.bib}

\end{document}